\documentclass[aps,twocolumn,showpacs,floatfix]{revtex4}
\usepackage{graphicx}
\usepackage{amsfonts}
\usepackage[figuresright]{rotating}  
\usepackage{amssymb}
\usepackage{amsmath}
\usepackage{psfrag}
\usepackage{subfigure}
\usepackage{multirow}
\usepackage{tabularx}
\usepackage{textcomp}
\usepackage{units}
\usepackage[active]{srcltx} 
\begin{document}

\title{Collective Motion of Polarized Dipolar Fermi Gases in the Hydrodynamic Regime}
\author{Aristeu R. P. Lima}
\email{lima@physik.fu-berlin.de}
\affiliation{Institut f\"{u}r Theoretische Physik, Freie Universit\"{a}t Berlin, Arnimallee 14, 14195 Berlin, Germany}
\author{Axel Pelster}
\email{axel.pelster@fu-berlin.de}
\affiliation{Fachbereich Physik, Universit\"{a}t Duisburg-Essen, Lotharstrasse 1, 47048 Duisburg, Germany}
\affiliation{Institut f\"{u}r Theoretische Physik, Freie Universit\"{a}t Berlin, Arnimallee 14, 14195 Berlin, Germany}
\begin{abstract}
Recently, a seminal STIRAP experiment 
allowed the creation of $^{40}$K$^{87}$Rb molecules in the 
rovibrational ground state [K.-K. Ni {\it et al.}, Science {\bf 322}, 231 (2008)]. 
In order to describe such a polarized dipolar Fermi gas in the hydrodynamic regime, 
we work out a variational time-dependent Hartree-Fock approach. 
With this we calculate dynamical properties of such a system as, for instance,
the frequencies of the low-lying excitations and the time-of-flight expansion. We find that the dipole-dipole interaction induces
anisotropic breathing oscillations in momentum space. In addition, after release from the trap, the momentum distribution becomes asymptotically isotropic, while the particle density becomes anisotropic.
\end{abstract}
\maketitle 

Even before the realization of Bose-Einstein condensation (BEC) with $^{52}$Cr \cite{PhysRevLett.94.160401}, much experimental and 
theoretical interest has been dedicated to ultracold quantum gases interacting through the long-range and anisotropic dipole-dipole interaction 
(DDI) \cite{review}. For bosonic dipolar particles, the starting point of the theoretical investigations was the construction 
of a corresponding pseudo-potential by Yi and You \cite{PhysRevA.61.041604}. After that, an exact solution of the Gross-Pitaevskii equation in the Thomas-Fermi 
regime was found for cylinder-symmetric traps \cite{PhysRevLett.92.250401}. Moreover, the DDI has been shown to shift the BEC critical 
temperature in a characteristic way in polarized systems \cite{glaum:080407} and to give rise to the Einstein-de-Haas effect, when spinorial 
degrees of freedom are considered \cite{kawaguchi:080405}. From the experimental point of view, time-of-flight (TOF) techniques demonstrated both 
the first DDI-signature through small mechanical effects \cite{PhysRevLett.95.150406} as well as strong dipolar effects in quantum 
ferrofluids \cite{strong-pfau}. Furthermore, the shape of the trap was manipulated to stabilize a purely dipolar BEC against 
collapse \cite{stabilization-pfau}.\\
Concerning fermionic dipolar  systems, recent theoretical studies have considered interesting properties of homogeneous gases such as zero 
sound \cite{ronen}, Berezinskii-Kosterlitz-Thoules phase transition \cite{bruun:245301}, and nematic 
phases \cite{fregoso}. In harmonically trapped systems, amazing predictions like anisotropic superfluidity \cite{baranov:250403}, 
fractional quantum Hall physics \cite{baranov:070404}, and Wigner crystallization \cite{baranov:200402} have been made. With respect to experimental 
investigations, the most promising atomic candidate is the fermionic chromium isotope $^{53}$Cr \cite{chicireanu:053406}, which has a magnetic moment 
of $m=6~$Bohr magnetons. For these atoms, calculations of equilibrium properties have shown that the DDI is only a small 
perturbation \cite{miyakawa:061603,zhang}. However, by applying a stimulated Raman adiabatic passage (STIRAP) process, it has recently been achieved to cool  
and trap 
$^{40}$K$^{87}$Rb molecules into their rovibrational ground-state, where they possess an electric dipole moment of $d=0.566$ 
Debye \cite{K.-K.Ni10102008,efficient,zirbel:013416,arXiv:0811.4618}. Due to the resulting strong DDI a considerable deformation of the momentum 
distribution is expected \cite{miyakawa:061603,zhang}. 
Once these systems would have been further cooled into the quantum degenerate regime, the main task will be to identify unambiguously the presence of the DDI. 
In this respect, TOF experiments and oscillation frequency measurements represent the most fundamental diagnostic tools in the field of ultracold  
quantum gases. Their outcomes reveal important information on the nature of the system under investigation. They differ drastically depending on whether the 
system is in the collisionless (CL) regime, where collision rates are small, or in the hydrodynamic (HD) regime, where collisions take place so often that they
lead to local equilibrium. To date, investigations of dynamical properties of trapped dipolar Fermi gases have either been  restricted to 
the CL regime \cite{1367-2630-11-5-055017} or excluded a deformation of the momentum distribution in the HD regime 
\cite{PhysRevA.67.025601}. Since the experiments 
with ultracold polar molecules are performed under strong dipolar interactions, one should expect 
them to lead the system into the HD regime, and thus an 
analysis allowing for an anisotropy in the momentum distribution has to be carried out. In this letter, we shall use a variational time-dependent 
Hartree-Fock approach to address this question.\\
Consider $N$ spin-polarized fermionic dipoles of mass $M$ trapped in a cylinder-symmetric harmonic potential 
$U_{\rm tr}({\mathbf x}) = {M}\omega^{2}_{x}\left( {x}^{2} + {y}^{2} + \lambda^{2}{z}^{2} \right)/{2}$ with trap anisotropy $\lambda$ at ultralow temperatures. Since the Pauli principle inhibits 
a contact interaction, they interact dominantly through DDI. As we assume that the fermionic cloud is polarized along the symmetry
axis of the trap, the DDI potential reads
$V_{\rm dd}({\mathbf x}) = \frac{C_{\rm dd}}{4\pi|{\mathbf x}|^{3}}\left[1-3\frac{z^{2}}{|{\mathbf x}|^{2}}\right].$
For magnetic dipole moments $m$ the DDI is characterized by $C_{\rm dd} = \mu_{0}m^{2}$, whereas for electric moments $d$ we 
have $C_{\rm dd} = 4\pi d^{2}$. In the following we restrict ourselves to the normal phase in the limit $T\rightarrow 0$ because the critical temperature for superfluidity is very low, depending exponentially on $a_{\rm dd}= MC_{\rm dd}/(4\pi\hbar^{2})$ \cite{baranov:250403}. Furthermore, this limit is restricted by the HD requirement that the relaxation time $\tau_{R}$ is small in comparison with the time scale $1/\overline{\omega}$ defined by the average trap frequency $\overline{\omega} = (\omega_{x}^{2} \omega_{z})^{1/3}$. The necessity of a HD approach can be inferred as follows. As $\tau_{R}$ is not kown for dipolar interactions, we estimate it by assuming the DDI to be equivalent to a contact interaction with scattering length $a_{\rm dd}$. Then we use the fact that for a two-component, degenerate, normal Fermi gas with contact interaction one has $(\overline{\omega}\tau_{R})^{-1}=(N^{1/3}a_{\rm dd}\sqrt{M\overline{\omega}/\hbar})^{2}F(T/T_{\rm F})$, where $F(T/T_{\rm F})$ is of the order 0.1 in the quantum temperature regime (see, e.g., \cite{vichi}). Thus, we expect for the one-component, dipolar gas to enter the HD regime for $N^{1/6}\epsilon_{\rm dd}\gg1$, with the dimensionless parameter $\epsilon_{\rm dd}={C_{\rm dd}} ({M^{3}\overline{\omega}}/{\hbar^{5}} )^{\frac{1}{2}}N^{\frac{1}{6}}/{4\pi}$ measuring the strength of the DDI. In the current set-up of Ref.~\cite{arXiv:0811.4618} one has $4\times10^{4}$ $^{40}$K$^{87}$Rb molecules  
with radial trapping frequency of ${\omega}_{x}=\omega_{y} \approx 2\pi\times 175$ Hz. Assuming an average trap frequency of that value yields at least $\epsilon_{\rm dd}\approx 5.3$ and $(\overline{\omega}\tau_{R})^{-1}\approx 0.1\times(N^{1/6}\epsilon_{\rm dd})^{2}\approx96$, which drives the system into the HD regime.

In this letter we work out a time-dependent Hartree-Fock approach by extremizing 
the action ${\cal A} = \int{\mathrm d}t \langle\Psi|i\hbar\frac{\partial}{\partial t} - \hat{H}|\Psi\rangle,$ where $\Psi(x_{1},\cdots,x_{N},t)=\langle x_{1},\cdots,x_{N}|\Psi\rangle$ 
is a Slater determinant and $\hat{H}$ denotes the underlying
Hamilton operator. In order to describe the HD regime, we follow a standard 
procedure of nuclear physics \cite{ring-schuck} and assume that frequent particle collisions assure that all one-particle orbitals have
the same local phase $\chi(x_{},t)$, yielding the velocity field ${\mathbf v}=\nabla\chi$. Thus, we can factorize out the phases and define a Slater determinant through 
$\Psi_{0}(x_{1},\cdots,x_{N},t) = e^{-iM \sum_{i}\chi(x_{i},t)/\hbar}\Psi(x_{1},\cdots,x_{N},t) $, which contains only the moduli of the one-particle orbitals and, therefore, is invariant under time reversal. This yields a time-even 
one-body density matrix 
$\rho_{0}(x,x';t)
=$ $e^{-iM\left[\chi(x,t)-\chi(x',t)\right]/\hbar}\rho_{}(x,x';t)$ \cite{brink}. With this 
the action reduces to
\begin{eqnarray}
{\cal A} & \!\!=\!\! &  -M\!\int\! {\mathrm d}t \!\!\!\int\!\! {\mathrm d}^{3}x\left\{ \dot{\chi}(x,t)\rho_{0}(x;t)
+\frac{\rho_{0}(x;t)}{2}\left[\nabla\chi(x,t)\right] ^{2} \right\}\nonumber\\
& & -  \int {\mathrm d}t\langle\Psi_{0}|\hat{H}|\Psi_{0}\rangle,
\label{action_rho0}
\end{eqnarray}
where $\rho_{0}(x;t)=\rho_{0}(x,x;t)$ denotes the particle density and $\langle\Psi_{0}|\hat{H}|
\Psi_{0}\rangle$
consists of the kinetic energy $E_{\rm ki}$, the trapping potential $E_{\rm tr}$, and the interaction. The latter
is divided into the direct or Hartree term $E_{\rm dd}^{\rm D}$ and the exchange or Fock term $E_{\rm dd}^{\rm E}$. Due to the exchange term, the 
ground-state energy $\langle\Psi_{0}|\hat{H}|\Psi_{0}\rangle$ is not a function of the particle density $\rho_{0}(x;t)$ alone, but also
contains the non-diagonal part $\rho_{0}(x,x';t)$. \\
\begin{figure}
\includegraphics[scale=.7]{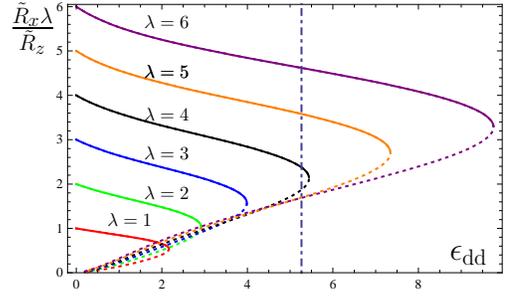}
\caption{(Color Online) Spatial aspect ratio for different trap anisotropies $\lambda$; 
the upper (continuous) branches correspond to a local minimum of the mean-field energy and the lower (dotted) branches to a maximum. 
Notice that the value of $\epsilon_{\rm dd}$ in which two branches meet, i.e., $\epsilon_{\rm dd}^{\rm crit}$, decreases slower for lower values of $\lambda_{}$. The vertical line marks the estimated value of the interaction strength for $^{40}$K$^{87}$Rb molecules $\epsilon_{\rm dd}\approx5.3$.}
\label{asp_rat_real_space}
\end{figure}
As it is not possible to solve analytically the resulting Euler-Lagrange equations for $\chi(x_{},t)$ and $\rho_{0}(x,x';t)$,
we propose here a 
variational extremization of the action. To this end, we express each energy contribution in terms of the Wigner transform of the one-body density 
matrix $\nu_{0}\left({\mathbf X},{\mathbf k};t\right) =$ 
$ \int {\mathrm d}^{3}s\,\rho_{0}\left({\mathbf X}+\frac{\mathbf s}{2},{\mathbf X}-\frac{\mathbf s}{2};t\right)\,e^{-i{\mathbf k}\cdot{\mathbf s}}$. 
The kinetic and trapping energy are then given by
\begin{eqnarray}
E_{\rm ki/tr} & = & \int \frac{{\mathrm d}^{3}x{\mathrm d}^{3}k} {(2\pi)^{3}}\,\nu_{0}\left({{\mathbf x}}{},{\mathbf k};t\right)
\epsilon_{\rm ki/tr}\left({{\mathbf x}}{},{\mathbf k}\right)
\label{E_kin}
\label{E_trap}
\end{eqnarray}
with $\epsilon_{\rm ki}=\hbar^{2}{\mathbf k}^{2} / 2M$ and $\epsilon_{\rm tr}=U_{\rm tr}({\mathbf x})$, 
respectively. The direct term, which accounts for the deformation of the particle density, and the exchange term, which is related to the momentum 
space deformation, read 
\begin{eqnarray}
E_{\rm dd}^{\rm D}\!\! & = &\!\! \int\!\! \frac{{\mathrm d}^{3}x{\mathrm d}^{3}k{\mathrm d}^{3}x'{\mathrm d}^{3}k'}{2(2\pi)^{6}} 
\nu_{0}\!\left({{\mathbf x}},{\mathbf k};t\right)\!V_{\rm dd}({\mathbf x}\!-\!{\mathbf x'})\nu_{0}\!\left({{\mathbf x'}},{\mathbf k'};t\right)\!, \nonumber\\
E_{\rm dd}^{\rm E}\!\! & = &\!\! -\!\!\int\!\! \frac{{\mathrm d}^{3}X{\mathrm d}^{3}k{\mathrm d}^{3}s{\mathrm d}^{3}k'}{2(2\pi)^{6}} 
\nu_{0}\!\left({{\mathbf X}},{\mathbf k};t\right)\!V_{\rm dd}({\mathbf s})\nu_{0}\!\left({{\mathbf X}},{\mathbf k'};t\right)\nonumber\\
\!\!\! & &\!\!\!\times e^{i{\mathbf s}\cdot({\mathbf k}-{\mathbf k'})} \, .
\end{eqnarray}
At this point, we adopt the variational ansatz $\chi(x,t) = \left[\alpha_{x}(t)(x_{}^{2} + y_{}^{2}) +\alpha_{z}(t)z_{}^{2}\right]/2$ for the phase
and
$\nu_{0}\left({{\mathbf x}},{\mathbf k};t\right) = \Theta\left(1-\frac{x^{2}+y^{2}}{R_{x}(t)^{2}}-\frac{z^{2}}{R_{z}(t)^{2}}
-\frac{k_{x}^{2}+k_{y}^{2}}{K_{x}(t)^{2}}-\frac{k_{z}^{2}}{K_{z}(t)^{2}} \right)$ for the Wigner phase space function
with $\Theta(\cdot)$ being the step function. We are now in the position to extremize the action (\ref{action_rho0}) with respect to the 
time-dependent variational parameters $\alpha_i (t)$ for the phase as well as $R_{i}(t)$ and $K_{i}(t)$ for the Thomas-Fermi radii and the Fermi momenta. 
At first, one obtains ${\alpha_{i}} = {\dot{R_{i}}}/{R_{i}}$, which is used to eliminate the parameters $\alpha_{i}$ 
from the rest of the formalism. Under conservation of the particle number
\begin{equation}
{\tilde{R}_{x}^{2} \tilde{R}_{z}} {\tilde{K}_{x}} ^{2} \tilde{K}_{z} = 1,
\label{part_num_cons}
\end{equation}
the equations of motion for the Thomas-Fermi radii read
\begin{eqnarray}
\hspace*{-2mm}\frac{1}{\omega_{x}^{2}}\frac{d^{2} {\tilde{R}}_{x}}{dt^{2}} \hspace*{-1mm} & = & \hspace*{-1mm}\!-\!\tilde{R}_{x} \!+\! \frac{2\tilde{K}_{x}^{2}\!
+\!\tilde{K}_{z}^{2}}{3\tilde{R}_{x}}\!+\!\epsilon_{\rm dd}A(\tilde{R}_{x},\tilde{R}_{z},\tilde{K}_{x},\tilde{K}_{z}),\\
\hspace*{-2mm}\frac{1}{\omega_{z}^{2}}\frac{d^{2}{\tilde{R}}_{z}}{dt^{2}} \hspace*{-1mm}& = & \hspace*{-1mm}\!-\!\tilde{R}_{z} \!+\! \frac{2\tilde{K}_{x}^{2}\!
+\!\tilde{K}_{z}^{2}}{3\tilde{R}_{z}}\!+\!\epsilon_{\rm dd}B(\tilde{R}_{x},\tilde{R}_{z},\tilde{K}_{x},\tilde{K}_{z}).
\label{eqn:dim_less_tf_eqs_cil}
\end{eqnarray}
Here we use $\tilde{\bullet}$ to represent the quantity $\bullet$ expressed in units of the non-interacting Thomas-Fermi radius
$R_{i}^{(0)} = \sqrt{{2E_{F}}/{M\omega_{i}^{2}}}$ and the Fermi momentum $K_{F} = \sqrt{{2E_{F}}/{\hbar^{2}}}$ with the Fermi energy 
$E_{F} =  \left(6N \right)^{{1}/{3}} \hbar \overline{\omega}$. The auxiliary functions are defined according to
\begin{eqnarray}
A & = & -\frac{c_{\rm d}}{\tilde{R}_{x}^{3}\tilde{R}_{z}}\left[ 1
- \frac{3\tilde{R}_{x}^{2}\lambda^{2}f_{s}\left({\tilde{R}_{x}^{}\lambda^{}}/{\tilde{R}_{z}}\right)}{2\left(\tilde{R}_{z}^{2}
-\tilde{R}_{x}^{2}\lambda^{2} \right)}  - f_{s}\left(\frac{\tilde{K}_{z}}{\tilde{K}_{x}}\right)\right],\nonumber\\
B & = & -\frac{c_{\rm d}}{\tilde{R}_{x}^{2}\tilde{R}_{z}^{2}}\left[ -2 + \frac{3\tilde{R}_{z}^{2}f_{s}\left({\tilde{R}_{x}^{}\lambda^{}} 
/{\tilde{R}_{z}}\right)} {\left(\tilde{R}_{z}^{2} - \tilde{R}_{x} ^{2} \lambda^{2} \right)}  - f_{s}\left(\frac{\tilde{K}_{z}}{\tilde{K}_{x}}\right)\right]
\nonumber
\label{AA_BB}
\end{eqnarray}
with the numerical constant $c_{\rm d}=\frac{2^{{38}/{3}}}{3^{{23}/{6}}\cdot5\cdot7\cdot\pi^{2}}\approx0.2791$.
\begin{figure}
\includegraphics[scale=.7]{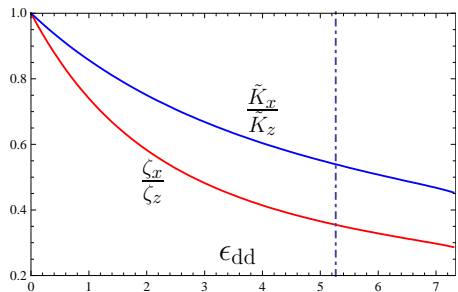}
\caption{(Color Online) The lower (red) curve shows the ratio of the amplitudes $\zeta_{x}/\zeta_{z}$ as a function of $\epsilon_{\rm dd}$ 
for $\lambda=5$. For comparison, the equilibrium aspect ratio in momentum space against $\epsilon_{\rm dd}$ for $\lambda=5$ is depicted by the upper (blue) curve.}
\label{zetaratio_fig}
\end{figure}
Furthermore, the anisotropy function 
\begin{eqnarray}
f_{s}(x) & \equiv & \begin{cases}
\frac{2x^2+1}{1-x^2}
- \frac{3 x^2\tanh^{-1}\sqrt{1-x^{2}}}{(1-x^{2})^{3/2}}; & x \neq 1\\
0; & x=1
\end{cases},
\end{eqnarray}
decreases monotonically from $1$ at $x=0$ to $-2$ at $x=\infty$, passing through zero at $x=1$ \cite{PhysRevLett.92.250401,glaum:080407}.
In addition, the variational parameters are restricted to obey
\begin{equation}
\tilde{K}_{z}^{2}  - \tilde{K}_{x}^{2}=  \epsilon_{\rm dd}C\left(\tilde{R}_{x},\tilde{R}_{z},\tilde{K}_{x}, \tilde{K}_{z}\right),
\label{constraint}
\end{equation}
with $C=\frac{3c_{\rm d}}{{\tilde{R}_{x}}^{2}\tilde{R}_{z}} \left[1 - \frac{\left(2\tilde{K}_{x}^{2} 
+ \tilde{K}_{z}^{2}\right)f_{s}\left({\tilde{K}_{z}}/{\tilde{K}_{x}} \right)}{ 2\left(\tilde{K}_{x}^{2} - \tilde{K}_{z}^{2} \right)} \right]$. 
This equation can be traced back to the exchange term and shows explicitly that a non-zero $\epsilon_{\rm dd}$ implies a deformed momentum 
distribution $\tilde{K}_{z}  \neq \tilde{K}_{x}$ for finite $\tilde{R}_{x}$, $\tilde{R}_{z}$ as was first pointed out in Ref. \cite{miyakawa:061603}. 
%

Equations (\ref{part_num_cons})--(\ref{eqn:dim_less_tf_eqs_cil}), (\ref{constraint}) govern the static as well as dynamic properties of a polarized
dipolar Fermi gas 
in the HD regime and represent the main result of this letter. They determine the temporal evolution of both the spatial and the
momentum distribution of a dipolar Fermi gas
which are directly experimentally accessible via TOF techniques. 
The static solutions agree precisely with the ones obtained before in Refs. \cite{miyakawa:061603,zhang}. 
In Fig. \ref{asp_rat_real_space} we present our findings for the spatial 
aspect ratio as a function of the dipolar strength $\epsilon_{\rm dd}$. The characteristic feature is that 
a minimal value of $\lambda$ is required for stabilizing a system with a 
given $\epsilon_{\rm dd}$. Thus, for future
experiments with $^{40}$K$^{87}$Rb molecules in the quantum degenerate regime one should choose the anisotropy $\lambda$ to be larger than the minimal value 
$\lambda_{\rm min}\approx 3.89$ in order to render the system stable against collapse. 
Amazingly, the minimum value of $\lambda$ supporting stability, decreases slowly and samples with $\lambda=0.05$ are stable if $\epsilon_{\rm dd}\lessapprox1.6$.
\begin{figure}
\includegraphics[scale=.7]{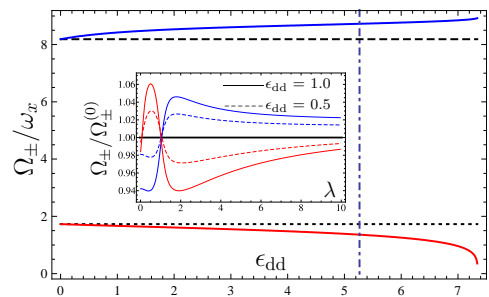}
\caption{(Color Online) Excitation frequencies for $\lambda=5$ as functions of the DDI-strength $\epsilon_{\rm dd}$. The upper blue (lower red) curve 
represents the monopole (quadrupole) frequency $\Omega_{+}$ ($\Omega_{-}$). The 
dashed (dotted) horizontal line represents the monopole (quadrupole) frequency of the non-interacting gas from Ref.~\cite{amoruso}. Inset: Mono- (blue) and quadrupole (red) oscillation frequencies of the dipolar Fermi gas normalized by the non-interacting values 
from Ref.~\cite{amoruso} against the trap aspect ratio $\lambda$ for different values of the dipolar strength $\epsilon_{\rm dd}$. The dashed (solid)
curves are for $\epsilon_{\rm dd}=0.5$ ($\epsilon_{\rm dd}=1.0$).}
\label{omegakrb}
\end{figure}

Having summarized
the most important aspects of the static solutions, we turn now to their dynamical properties. In a cylinder-symmetric system the mono- and quadrupole low-lying oscillation modes couple to each other. In order to obtain the frequency of these 
modes in the HD regime, we expand the radii and momenta around their respective equilibrium values according to $\tilde{R}_{i} = \tilde{R}_{i}{(0)} 
+ \eta_{i}e^{i\Omega t}, \tilde{K}_{i} = \tilde{K}_{i}{(0)} + \zeta_{i}e^{i\Omega t}$, where $\eta_{i}$ ($\zeta_{i}$) denotes a small oscillation amplitude 
in the $i$-th direction in real (momentum) space and $\Omega$ represents the oscillation frequency. Inserting these into the equations of motion 
(\ref{part_num_cons})--(\ref{eqn:dim_less_tf_eqs_cil}), (\ref{constraint}), a linearization yields at first
for the ratio of the momentum amplitudes 
\begin{equation}
\frac{\zeta_{x}}{\zeta_{z}} = \frac{\tilde{K}_{x}}{\tilde{K}_{z}}\frac{\tilde{K}_{x}^{2}+\tilde{K}_{z}^{2}-\epsilon_{\rm dd}\tilde{K}_{z}
\partial C / \partial \tilde{K}_{z}}{2\tilde{K}_{z}^{2}-\epsilon_{\rm dd}\tilde{K}_{z} \partial C / \partial \tilde{K}_{z}}\,,
\end{equation}
where all terms are evaluated at equilibrium. This quantity is plotted against $\epsilon_{\rm dd}$ for $\lambda=5$ in the
red (lower) curve in Fig.~\ref{zetaratio_fig}  and is compared
to the corresponding equilibrium momentum aspect ratio (blue, upper curve). Setting $C=0$, i.e., removing the exchange term, one has 
${\zeta_{x}}={\zeta_{z}}$, whereas for non-zero $C$, the ratio ${\zeta_{x}}/{\zeta_{z}}$ decreases 
monotonically from $1$ to about $0.28$ in the interval $0<\epsilon_{\rm dd}<\epsilon_{\rm dd}^{\rm crit}\approx 7.34$. This shows that the exchange term induces characteristic {\it anisotropic breathing oscillations 
in mo\-men\-tum space}, which can be regarded as a tra\-de\-mark sign of the DDI in fer\-mi\-onic quantum gases.\\
Eliminating the momentum amplitudes $\zeta_i$ yields a reduced linear homogeneous system for the spatial amplitudes $\eta_i$. 
Demanding non-trivial solutions  yields an explicit but lengthy result for
the monopole (quadrupole) oscillation frequency $\Omega_+$ ($\Omega_-$) which depends via the equilibrium values of the Thomas-Fermi radii 
and the Fermi momenta upon the trap anharmonicity $\lambda$ and the
dipolar strength $\epsilon_{\rm dd}$. In the special case of an ideal Fermi gas, i.e. 
$\epsilon_{\rm dd}=0$, the oscillation frequencies $\Omega_{\pm}$ reduce to the correct non-interacting values
${{\Omega_{\pm}^{(0)}}^{2}} = {\omega_{x}^{2}}\left({5+4\lambda_{}^{2}} \pm \sqrt{25-32\lambda_{}^{2} + 16\lambda_{}^{4}}\right)/3$, which were first 
obtained for $\lambda=1$ in Ref.~\cite{PhysRevLett.83.5415} and for $\lambda\neq1$ in Ref.~\cite{amoruso}. Fig.~\ref{omegakrb} shows the oscillation frequencies of the mono- (blue) and quadrupole (red) modes plotted against 
$\epsilon_{\rm dd}$ for $\lambda=5$.
As $\epsilon_{\rm dd}$ becomes larger, we find that the monopole frequency increases and that the quadrupole frequency decreases, vanishing at $\epsilon_{\rm dd}^{\rm crit}\approx 7.34$, the same value for which the system becomes unstable (see Fig.~\ref{asp_rat_real_space}). 
The inset of  Fig.~\ref{omegakrb} shows how the frequencies depend on the anisotropy $\lambda$  for $\epsilon_{\rm dd} = 0.5$ (dashed) 
and $\epsilon_{\rm dd} = 1.0$ (continuous). It turns out that the quadrupole frequencies are larger than in the non-interacting case for $\lambda<1$ 
and smaller for $\lambda>1$, while the contrary is true for the monopole modes. This behaviour agrees qualitatively with dipolar BECs 
\cite{PhysRevLett.92.250401}.\\
\begin{figure}[t]
\includegraphics[scale=.65]{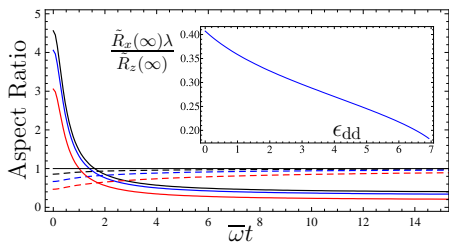}
\caption{(Color Online) Cloud aspect ratio in TOF expansion for $\lambda=5$ with $\epsilon_{\rm dd}=1,3$ and $7$ (continuous, top to bottom). The dashed curves depict the corresponding momentum aspect ratios. Inset: asymptotic cloud aspect ratio against $\epsilon_{\rm dd}$.}
\label{tof}
\end{figure}
It remains to study the TOF expansion of a dipolar Fermi gas. This is done by numerically solving the 
Eqs.~(\ref{part_num_cons})--(\ref{eqn:dim_less_tf_eqs_cil}), (\ref{constraint}), while removing the trap frequencies. The 
results are presented in Fig.~\ref{tof}, where the spatial and momentum aspect ratios are plotted as functions of time in units of $\overline{\omega}^{-1}$ 
at $\lambda=5$ for different  $\epsilon_{\rm dd}$. 
The characteristic of the hydrodynamic regime is that the asymptotic value of the aspect ratio in real space depends on $\epsilon_{\rm dd}$, while local equilibrium renders the momentum distribution asymptotically isotropic. We can estimate the validity of these results if we assume the previous HD criterion to be valid also during the expansion. Since the equations of motion imply $d^{2}{\tilde{R}}_{i}/d t^{2}=0$ for large times, yielding $(\tilde{R_{x}^{2}}\tilde{R_{z}})^{1/3}\sim\overline{\omega}t$, one obtains a HD expansion provided $(\overline{\omega}t)^{2}\cdot\overline{\omega}\tau_{R}\ll 1$. For KRb molecules, the expansion is HD only for small times $\overline{\omega}t\ll10$, whereas for molecules like LiCs with $d\approx5.5~$Debye, the expansion is HD for $\overline{\omega}t\ll1000$.

In the present letter we have investigated both low-lying oscillation frequencies and TOF expansion data for a polarized dipolar Fermi gas through a hydrodynamic approach. Our findings have revealed different fingerprints of a strong DDI. We have estimated the validity of our results and found strong evidence for hydrodynamic behavior also in the absence of superfluidity. The prospects for observing normal dipolar hydrodynamics in the quantum degenerate regime are enhanced by tight traps and the recently obtained large dipole moments.
\\
We acknowledge support from the DAAD, the Innovationsfond FU-Berlin, and from the DFG in SFB/TR12.

\end{document}